\documentclass[aps,prl,twocolumn,showpacs,amssymb,amsmath]{revtex4}
\usepackage{graphicx}
\usepackage{dcolumn}
\usepackage[colorlinks=false,dvipdfm]{hyperref} 

\begin{document}

\title{Observation of Geometric Phases for Three-Level Systems using NMR Interferometry}

\author{Hongwei Chen$^{1}$,Mingguang Hu$^{2}$,Jingling Chen$^{2}$,Jiangfeng Du$^{1}$}
\email{djf@ustc.edu.cn}

\affiliation{$^1$Hefei National Laboratory for Physical Sciences at
Microscale and Department of Modern Physics, University of Science
and Technology of China, Hefei, Anhui 230026,
People's Republic of China\\
$^2$Theoretical Physics Division, Chern Institute of Mathematics,
Nankai University, Tianjin 300071, China }

\date{\today}

\begin{abstract}
Geometric phase (GP) independent of energy and time rely only on the
geometry of state space. It has been argued to have potential fault
tolerance and plays an important role in quantum information and
quantum computation. We present the first experiment for producing
and measuring an Abelian geometric phase shift in a three-level
system by using NMR interferometry. In contrast to existing
experiments, based on the geometry of $S^2$, our experiment concerns
the geometric phase with the geometry of $SU(3)/U(2)$. Two
interacting qubits have been used to provide such a
three-dimensional Hilbert space.
\end{abstract}


\maketitle

When a quantum mechanical system evolves cyclically in time so that
it returns to its initial physical state, its wave function can
acquire a geometric phase factor in addition to the familiar dynamic
phase \cite{1992Anandan}. If the cyclic change of the system is
adiabatic, this additional factor is known as Berry's phase
\cite{1984Berry}. Otherwise, it is related to Aharonov-Anandan (AA)
phase \cite{1987Aharonov} that has been pointed out to be a
continuous version of earlier Pancharatnam phase \cite{1956Panch}.

Geometric phases (GP) independent of energy and time rely only on
the geometry of state space. It is therefore resilient to certain
types of errors and suggests the possibility of an intrinsically
fault-tolerant way of performing quantum gate operations
\cite{2000Nielson,2000Jones,2003Leibfried}. This potential value
makes it important to observe and further apply GP in different
quantum physical systems. The observations of GP began from earlier
spin-polarized neutrons through a solenoid \cite{1987Bitter},
polarized light through a helically twisted optical fibre
\cite{1986Tomita}, and a pair of coupled protons in magnetic field
using NMR \cite{1989Suter} to the recent superconducting qubit
experiment \cite{2007Leek}. The principle of them are usually the
same. That is, in a two-level state space the geometry of it
corresponds to a sphere $S^2$ and the GP equals to one half the
solid angle subtended by closed paths on $S^2$.

When one generalizes to a three-level quantum system
\cite{1997Khanna,1997Arvind}, the geometry of $S^2$ gets replaced by
a four-dimensional geometric space $SU(3)/U(2)$ or part of sphere
$S^7$. Then evolutions of state correspond to actions of $SU(3)$ on
$SU(3)/U(2)$ that is different from that of $SU(2)$ on $S^2$ for the
two-level case. In order to observe GP, one way to vanish the
dynamical phase is closely linked to the geodesic in ray space (see
below). For two-level case, the geodesic in ray space happens to
coincide with that on $S^2$. In contrast, it is a plane curve
instead of geodesic on $S^7$ for three-level case. The GP for any
cyclic evolution in three-level ray space are no longer related to
solid angles on $S^7$ but referred to Bargmann invariants
\cite{1964Bargmann,1993Mukunda}. All of these differences indicate
the observation of three-level GP technically more difficult
\cite{2001Sander}.

In this letter, we report an experimental observation of three-level
GP by using NMR interferometry. The three levels referred in the
experiment are chosen from a two spin-$1/2$ interacting system.
Unitary evolutions for implementing cyclic paths in the
three-dimensional ray space are ensured by quantum controlled logic
gate operations \cite{2004Vandersypen}. Aimed at obtaining a
measurable GP, we evolve the target state while keeping the
reference state unchanged to produce a relative phase between them.
\begin{figure}[tbph]
\centering
\includegraphics[width=0.8\columnwidth]{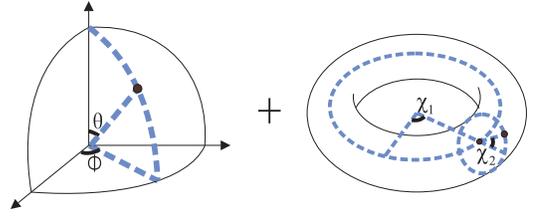}
\caption{An illustration of the parameter space for all three-level
states. The local coordinates $\theta$, $\phi$, $\chi_1$ and
$\chi_2$ are such that $(\theta,\phi)$ define a point in the
positive octant of $S^2$ and $(\chi_1,\chi_2)$ for $(\theta,\phi)$
fixed define a point on a torus.} \label{fig1}
\end{figure}

In the three-dimensional Hilbert space $\mathcal{H}^{3}$, an
arbitrary state can be expressed as
\begin{equation}\label{eq-psi}
|\psi\rangle=e^{i\eta}\big(e^{i\chi_1}\cos\theta,e^{i\chi_2}\sin\theta\cos\phi,\sin\theta\sin\phi\big),
\end{equation}
where the real parameters have the range
$\theta,\phi\in[0,\frac{\pi}{2}]$ and $\chi_1,\chi_2\in[0,2\pi)$. It
has an one-to-one correspondence, omitting a global phase $\eta$, to
the point on an octant of $S^2$ plus a torus (Fig. \ref{fig1}).
Consider a state evolves from $|\psi(s_1)\rangle$ to
$|\psi(s_2)\rangle$ with $s_{1,2}$ being the curve parameters
determined by the Eq. (\ref{eq-psi}). Corresponding to this
evolution, in $\mathcal{H}^3$ there is a continuous piecewise smooth
parametrized curve, $C=\big\{\psi(s)\big|s_1\leq s\leq s_2\big\}$,
and its image in the ray space $\mathcal{R}$ is likewise continuous
and piecewise smooth denoted by
$\mathcal{C}=\big\{\rho(s)=|\psi(s)\rangle\langle\psi(s)|\big|s_1\leq
s \leq s_2\big\}$. Then the GP $\beta$ associated with the cure
$\mathcal{C}$ equals to the difference between a total
phase $\varphi_{\text{tot}}$ and a dynamical phase
$\gamma_d$ \cite{1997Arvind}, that is,
\begin{eqnarray}
\beta[\mathcal{C}]&=&\varphi_{\text{tot}}[C]-\gamma_d[C],\nonumber\\
\varphi_{\text{tot}}[C]&=&\mathrm{arg}\langle \psi(s_1)|\psi(s_2)\rangle,\label{eq-phase}\\
\gamma_d[C]&=&-\int^{s_2}_{s_1}ds
\langle\psi(s)|i\frac{\partial}{\partial s}|\psi(s)\rangle,\nonumber
\end{eqnarray}
with both $\varphi_{\text{tot}}$ and $\gamma_d$ being functionals of
the curve $C$. If the curve $\mathcal{C}$ is closed, the state
change can be simply expressed as
$|\psi(s_2)\rangle=\exp\{i(\gamma_d[C]+\beta[\mathcal{C}])\}|\psi(s_1)\rangle$.


The geodesics in ray space $\mathcal{R}$  are given through
variations of a nondegenerate positive definite length functional
\cite{1993Mukunda}.
In two-level systems geodesics are related to the parellel transport
condition. For the three-level case,  every geodesic in ray space
has a vanishing geometric phase and it plays a crucial role in the
observation of geometric phases in the following.
The simplest description of geodesic can always be achieved as
follows \cite{1997Arvind}. Let $\rho_k$ and $\rho_{k+1}$ denote the
end points of a smooth curve $\mathcal{C}$ associated with unit
vectors $\psi_k$ and $\psi_{k+1}$ in $\mathcal{H}^{3}$. There is a
requirement for the chosen state vectors that
$\langle\psi_k|\psi_{k+1}\rangle$ must be real positive. Then the
geodesic $\mathcal{C}_{\text{geo}}$ connecting $\rho_k$ to
$\rho_{k+1}$ is the ray space image of the curve
$C_{\text{geo}}=\{\psi(s_k)|0\leq s_k\leq s_k^0\}$ and
\begin{equation}\label{eq-geo}
\psi(s_k)=\psi_k\cos
s_k+\frac{\psi_{k+1}-\psi_k\langle\psi_{k}|\psi_{k+1}\rangle}{\sqrt{1-\langle\psi_k|\psi_{k+1}\rangle^2}}\sin
s_k,
\end{equation}
with $0\leq s_k\leq s_k^0$ and
$s_k^0=\arccos\langle\psi_{k+1}|\psi_k\rangle$. From the Eq.
(\ref{eq-geo}), one can see that $\psi(0)=\psi_k$ and
$|\psi(s_k^0)\rangle\langle\psi(s_k^0)|=|\psi_{k+1}\rangle\langle\psi_{k+1}|=\rho_{k+1}$.
For a set of points $\rho_1,\rho_2,\cdots,\rho_n\subset \mathcal{R}$
in order, suppose that no two consecutive points are mutually
orthogonal and that $\rho_n$ and $\rho_1$ are also nonorthogonal. So
we can obtain a closed curve $\mathcal{C}$ in $\mathcal{R}$ in the
form of a $n$-sided polygon by joining these $n$ points cyclically
with geodesic arcs. The geometric phase is then according to the Eq.
(\ref{eq-phase})
\begin{eqnarray}
\beta[\mathcal{C}]&=&\mathrm{arg}\langle\psi_1|\psi_1'\rangle-\mathrm{arg}\langle\psi_1|\psi_2\rangle-\cdots-\mathrm{arg}\langle\psi_n|\psi_1'\rangle\nonumber\\
&=&-\mathrm{arg}\mathrm{Tr}(\rho_1\rho_2\cdots\rho_n),\label{eq-phg}
\end{eqnarray}
in which it has used relations of
$|\psi_1'\rangle\langle\psi_1'|=|\psi_1\rangle\langle\psi_1|=\rho_1$
and $\rho_1^2=\rho_1$. The Eq. (\ref{eq-phg}) combined with geodesic
condition, i.e., $\langle\psi_k|\psi_{k+1}\rangle$ is real positive,
shows a vanishing dynamical phase for these cyclic evolutions. It
thus provides us a convenient evolution way to observe the geometric
phase.

\begin{figure}
  \includegraphics[width=0.7\columnwidth]{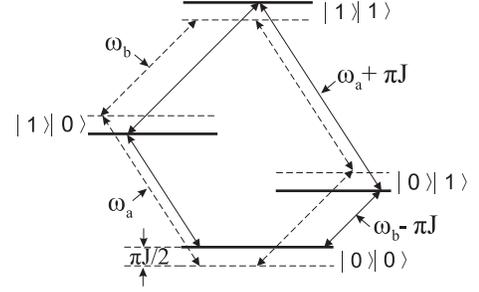}\\
  \caption{Energy level diagram for (solid lines) two spins coupled by a Hamiltonian of the
form of $2\pi\hbar JI_z^1I_z^2$  and (dashed lines) two uncoupled
spins.}\label{fig2}
\end{figure}

Experiments are performed on the three-dimensional subspace of two
interacting spin-$\frac{1}{2}$ nuclei---spin $a$ ($^{13}$H) and spin
$b$ ($^{1}$C) in the $^{13}$C-labeled chloroform molecule
$\mathrm{CHCl_3}$. The reduced Hamiltonian for this two spin system
is, to an excellent approximation, given by $H = \omega_aI^a_z +
\omega_b I^b_z + 2\pi J I^a_zI^b_z $. The first two terms in the
Hamiltonian decribe the free precession of spin \emph{a} and spin
\emph{b} around the magnetic field $B_0$ with Larmour frequencies
$\omega_a/2\pi\approx400$ MHz and $\omega_b/2\pi\approx100$ MHz. The
third term of the Hamiltonian describes a scalar spin-spin coupling
of the two spins with $J=214.5$ Hz. Experiments were performed at
room temperature on a Bruker AV-400 spectrometer. If we denote the
spin up and down by $|0\rangle$ and $|1\rangle$, the energy levels
of such system are displayed in Fig. \ref{fig2}. It has four levels
written as $\{|00\rangle,|01\rangle,|10\rangle,|11\rangle\}$
corresponding to energy eigenvalues
$\{\frac{1}{2}\hbar(-\omega_1-\omega_2+\pi
J),\frac{1}{2}\hbar(-\omega_1+\omega_2-\pi
J),\frac{1}{2}\hbar(\omega_1-\omega_2-\pi
J),\frac{1}{2}\hbar(\omega_1+\omega_2+\pi J)\}$. We choose basis
states $\{|00\rangle$, $|10\rangle$, $|11\rangle\}$ to construct the
desired three-level space $\mathcal{H}^3$ and $|01\rangle$ as the
reference state which keeps unchanged during evolutions.

The system was first prepared in a pseudopure state $|00\rangle$
using the method of spatial averaging \cite{Cory120} with the pulse
sequence
\begin{equation}\label{inintial}
R^b_x (\pi/3 ) \rightarrow G_z \rightarrow R^b_x (\pi/4 )
\rightarrow \frac{1}{2J} \rightarrow R^b_y( \pi/4 ) \rightarrow G_z,
\end{equation}
which is read from left to right (as the following sequences). The
rotations  $R^{spins}_{axis}(angle)$ are implemented by
radio-frequency pulses. $G_z$ is a pulsed field gradient which
destroys all coherences (x and y magnetizations) and retains
longitudinal magnetization (z magnetization component) only. $
\frac{1}{2J}$ represents a free precession period of the specified
duration under the coupling Hamiltonian (no resonance offsets).

\begin{figure}[tbph]
  \includegraphics[width=1\columnwidth]{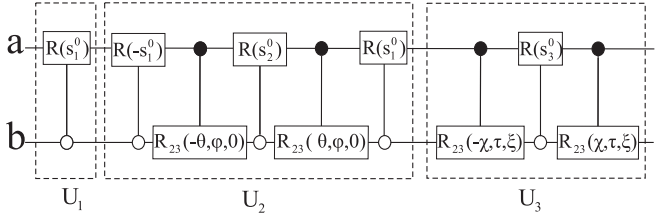}\\
  \caption{Experimental network: two spin-$1/2$ nuclei perform unitary
evolutions controlled by each other. Each circle at the second line
means that performs its linked unitary evolution when the second
nucleus at $|0\rangle$ state. Each dot at the first line means that
performs its linked unitary evolution  when the first nucleus at
$|1\rangle$ state.}\label{fig3}
\end{figure}

The complete sequence started by preparing the initial superposition
state $\frac{1}{\sqrt{2}}(|00\rangle+|01\rangle)$ with a Hadamard
operation on the second qubit of the pseudopure state $|00\rangle$.
Then the reference term $|01\rangle$ was kept unchanged through
bipartite control operations as shown in Fig. \ref{fig3}. The
$|00\rangle$ term (denoted by $|\psi_1\rangle$) was first evolved to
$|\psi_2\rangle=\cos s_1^0|00\rangle+\sin s_1^0|10\rangle$ with
unitary operation $U_1^\text{g}(s_1)$, then to state
$|\psi_3\rangle=(\cos s_1^0\cos s_2^0-e^{i\theta}\sin s_1^0\sin
s_2^0\cos\varphi)|00\rangle+(\sin s_1^0\cos s_2^0+e^{i\theta}\cos
s_1^0\sin s_2^0\cos\varphi)|10\rangle+\sin\varphi \sin
s_2^0|11\rangle$ with $U_2^\text{g}(s_2)$, and last to state
$|\psi_1'\rangle=e^{i\beta}|\psi_1\rangle$ with $U_3^\text{g}(s_3)$.
Corresponding to three smooth geodesics, the unitary operations can
be factored into more clear form
\begin{eqnarray}
U_1^{\text{g}}(s_1)&=&R(s_1),\nonumber\\
U_2^{\text{g}}(s_2)&=&R(s_1^0)R_{23}(\theta,\varphi,0)\
R(s_2)\  R_{23}^{-1}(\theta,\varphi,0)R^{-1}(s_1^0),\nonumber\\
U_3^{\text{g}}(s_3)&=&R_{23}(\chi,\tau,-\xi)\ R(-s_3)\
R_{23}^{-1}(\chi,\tau,-\xi),\label{eq-u}
\end{eqnarray}
where
\begin{equation*}
R(s_k)=\begin{pmatrix} \cos s_k&0&-\sin s_k&0\\
0&1&0&0\\
\sin s_k&0&\cos s_k&0\\
0&0&0&1
\end{pmatrix},
\end{equation*}
and the $SU(2)_{23}$ subgroup element
\begin{equation*}
R_{23}(\theta,\phi,\varphi)=\begin{pmatrix} 1&0&0&0\\
0&1&0&0\\
0&0&e^{i\phi}\cos \theta&e^{-i\varphi}\sin\theta\\
0&0&-e^{i\varphi}\sin\theta&e^{-i\phi}\cos\theta
\end{pmatrix}.
\end{equation*}
Parameters $\xi,\chi,\tau,s_3^0$ in Eq. (\ref{eq-u}) are determined
by the reparametrization $|\psi_3\rangle=e^{i\xi}\cos
s_3^0|00\rangle+e^{i(\xi+\chi)}\sin s_3^0\cos\tau|10\rangle+\sin
s_3^0\sin\tau|11\rangle$ and curve parameters $s_k$ ($k=1,2,3$) have
ranges $0\leq s_k\leq s_k^0$ .
Obviously the chosen unit vectors $\psi_k$ and $\psi_{k+1}$ satisfy
the condition of $\langle\psi_k|\psi_{k+1}\rangle$ being real
positive. This combined with the Eq. (\ref{eq-phg}) shows a
vanishing dynamical phase during these cyclic evolutions and we
obtain the GP
\begin{equation}\label{eq-beta}
\beta[\mathcal{C}]=\mathrm{arg}(\cos s_1^0\cos s_2^0-e^{i\theta}\sin
s_1^0\sin s_2^0\cos\varphi).
\end{equation}
So after one cyclic evolution described above, it effectively
produces a GP and can be measured as a relative phase shift between
$|0\rangle_b$ and $|1\rangle_b$ for the qubit $b$, i.e.,
$\frac{1}{\sqrt{2}}(|00\rangle+|01\rangle)\rightarrow\frac{1}{\sqrt{2}}
(e^{i\beta}|00\rangle+|01\rangle)\rightarrow|0\rangle_a\otimes\frac{1}{\sqrt{2}}(e^{i\beta}|0\rangle+|1\rangle)_b$.
At last the local phase $\beta$ can be read out directly by a phase
sensitive detector on qubit $b$ in NMR.

\begin{figure}
\includegraphics[width=1\columnwidth]{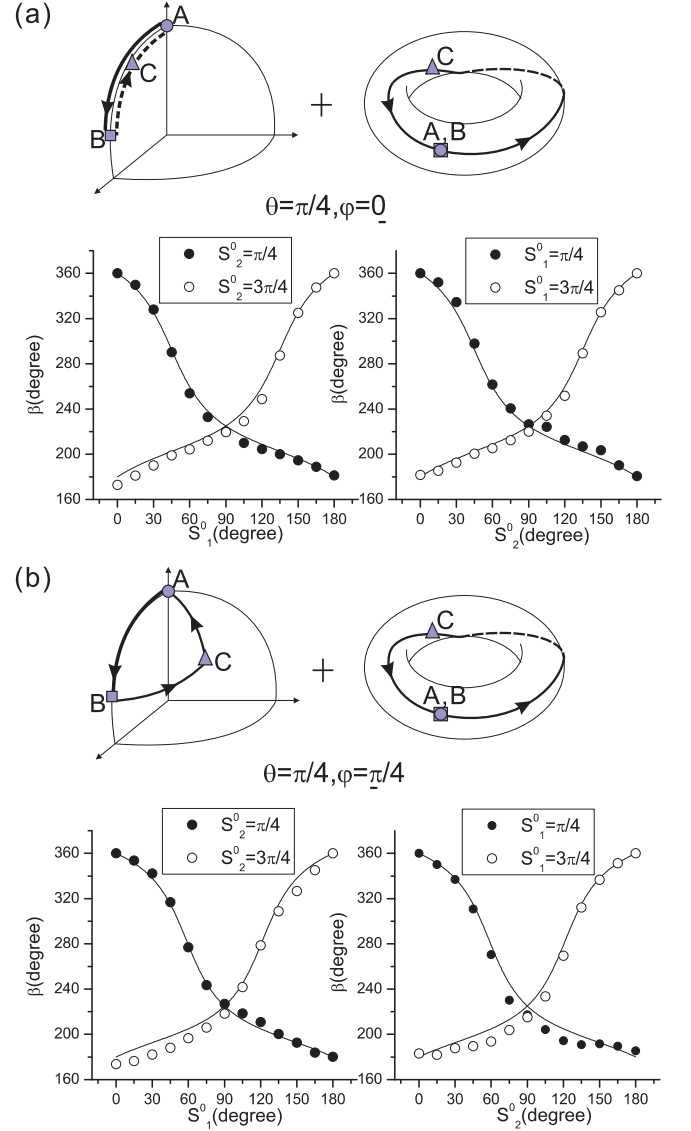}\\
\caption{Experimental results on the GP $\beta$ versus the parameter
$s_1^0$ or $s_2^0$. Changing $s_1^0$, $s_2^0$ means changing the
positions of the points $B$ and $C$. The evolution paths have been
depicted out in parameter space and the theoretical curves are
marked out by lines. (a) It shows the result in the case of
$\theta=\pi/4$ and $\varphi=0$; the $ABC$ on the octant of $S^2$ is
a curve while it runs a period on the torus. (b) It shows the result
in the case of $\theta=\pi/4$ and $\varphi=\pi/4$; the $ABC$ on the
octant of $S^2$ is a triangle  while it runs a period on the torus.
}\label{fig4}
\end{figure}

In Fig. \ref{fig4} we show the measured phase $\beta$ and its
dependence on different points $A,B,C$ in ray space characterized by
parameters $\{s_1^0,s_2^0,\theta,\varphi\}$, all carried out at
$\theta=\pi/4$, and total pulse sequence time $T$ for cyclic
evolution is about $5\sim25ms$ for different evolution path. In Fig.
\ref{fig4} (a) and (b), we set $\varphi=0$ and $\varphi=\pi/4$
respectively with which geodesics have disparate trajectories in ray
space. The measured phase is in all cases seen to fit the
theoretical curve (\ref{eq-beta}) well with a root-mean-square
deviation across all data sets of $5.6$ degree. Thus, all results
are in close agreement with the predicted geometric phase, and it is
clear that we are able to accurately control the amount of phase
geometrically.

The controlled operation between the two qubits plays the main role
in the experiment. It goes as that the qubit $a$ (or $b$) undergoes
a $SU(2)$ operation if the qubit $b$ (or $a$) is in $|1\rangle$
while kept unchanged if it in $|0\rangle$. This is used to realize
the controlled operations $R$ and $R_{23}$. The concrete operations
go as follows.

For the subsystem of qubit a, we can write the reduced Hamiltonian
$$
H_a =   \omega_a I^a_z + 2 \pi J m_z^b I^a_z
 =  [ \omega_a - 2 \pi J (d^b -\frac{1}{2})] I^a_z   ,
$$
where $m_z^b$ is the eigenvalue of $I_z^b$ ($=\pm\frac{1}{2}$) and
$d^b$ the corresponding computational value ($=0,\,1$). If we use a
rotating frame with a frequency of $\omega_{a}'=\omega_{a}$ and
$\omega_{b}'=\omega_{b}$, the Hamiltonian turns into, for $d^b=0$,
$H_a^{(0)}=\pi J I^{a}_{z}$, while it becomes $H_{a}^{(1)}=-\pi J
I^{a}_{z}$ for $d^b=1$. This Hamiltonian generates controlled
rotations around the z-axis. To generate the control gate $R(S_k)$,
we rotate the rotation axis using radio-frequency pulses. To
generate a $2S_k$ rotation around the $y$-axis, e.g., we use the
sequence
$$
R_{x}^{a}(\frac{\pi}{2})\rightarrow \frac{S_k}{\pi J}\rightarrow
R_{x}^{a}(-\frac{\pi}{2})\rightarrow R_{y}^{a}(S_k) .
$$
This represents the controlled gate operation $R(S_k)$.

For another controlled gate operation $R_{23}(\chi,\tau,-\xi)$, we
have to reverse the roles of control and target qubit and apply the
following sequence to qubit b:
\begin{eqnarray*}
&R_{z}^{2}(-\phi) \rightarrow R_{y}^{2}(-\pi-\beta) \rightarrow
\frac{\alpha}{2\pi J} \rightarrow R_{y}^{2}(\pi+\beta)&\nonumber\\
&\rightarrow R_{z}^{2}(\phi) \rightarrow
R_{n}^{2}(\alpha,\beta,\phi),& \label{e:U2}
\end{eqnarray*}
$R_{n}^{2}(\alpha,\beta,\phi)$ denote to rotate the second qubit
$\alpha$ around the axis $\vec{n}(\beta,\phi)$, and
$\alpha,\beta,\phi$ is calculated from $\chi,\tau,-\xi$.



In conclusion, when a quantum mechanical system evolves cyclically
in time so that it returns to its initial physical state, its wave
function can acquire a geometric phase factor in addition to the
familiar dynamic phase. Geometric phases (GP) independent of energy
and time rely only on the geometry of state space. It is therefore
resilient to certain types of errors and suggests the possibility of
an intrinsically fault-tolerant way of performing quantum gate
operations. we present the first experiment for producing and
measuring an Abelian geometric phase shift in a three-level system
by using NMR interferometry. In contrast to existing experiments,
based on the geometry of $S^2$, our experiment concerns the
geometric phase with the geometry of $SU(3)/U(2)$. Two interacting
qubits have been used to provide such a three-dimensional Hilbert
space.

We would like to thank Prof. Zhang for inspiring conversations. This
work was supported by the National Natural Science Foundation of
China, the CAS, Ministry of Education of PRC, and the National
Fundamental Research Program. This work was also supported by
European Commission under Contact No. 007065 (Marie Curie
Fellowship). J.-L. C. acknowledges supports in part by NSF of China
(Grant No. 10575053 and No. 10605013) and Program for New Century
Excellent Talents in University.


\begin{thebibliography}{30}
\bibitem{1992Anandan}
J. Anandan, \emph{The geometric phase.} Nature \textbf{360}, 307-313
(1992).
\bibitem{1984Berry}
M. V. Berry, Proc. R. Soc. \textbf{A392}, 45-57 (1984).
\bibitem{1987Aharonov}
Y. Aharonov and J. Anandan, \emph{Phase Change during a Cyclic
Quantum Evolution.} Phys. Rev. Lett. \textbf{58}, 1593-1596 (1987).
\bibitem{1956Panch}
S. Pancharatnam, Proc. Indian Acad. Sci. \textbf{A44}, 247 (1956).
\bibitem{2000Nielson}
M. A. Nielson and I. L. Chuang, \emph{Quantum computing and Quantum
Information} (Cambridge Univ. Press, Cambridge, 2000).
\bibitem{2000Jones}
J. A. Jones, V. Vedral, A. Ekert, and G. Castagnoli, \emph{Geometric
quantum computation using nuclear magnetic resonance.} Nature
\textbf{403}, 869-871 (2000).
\bibitem{2003Leibfried}
D. Leibfried et al., \emph{Experimental demonstration of a
robust,high-fidelity geometric two ion-qubit phase gate.} Nature
\textbf{422}, 412-415 (2003).
\bibitem{1987Bitter}
T. Bitter and D. Dubbers, \emph{Manifestation of Berry’s
topological phase in neutron spin rotation.} Phys. Rev. Lett.
\textbf{59}, 251-254 (1987).
\bibitem{1986Tomita}
A. Tomita and R. Y. Chiao, \emph{Observation of Berry's Topological
Phase by Use of an Optical Fiber.} Phys. Rev. Lett. \textbf{57},
937-940 (1986).
\bibitem{1989Suter}
D. Suter, K. T. Mueller, and A. Pines, \emph{Study of the
Aharonov-Anandan Quantum Phase by NMR Interferometry.} Phys. Rev.
Lett. \textbf{60}, 1218 (1988).
\bibitem{2007Leek}
P. J. Leek et al., \emph{Observation of Berry's Phase in a
Solid-State Qubit.} Science \textbf{318}, 1889-1892 (2007).
\bibitem{2004Vandersypen}
L. M. K. Vandersypen and I. L. Chuang, \emph{NMR techniques for
quantum control and computation.} Rev. Mod. Phys. \textbf{76}, 1037
(2004).

\bibitem{1997Khanna}
G. Khanna, S. Mukhopadhyay, R. Simon, and N. Mukunda,
\emph{Geometric Phases for $SU(3)$ Representations and Three Level
Quantum Systems.} Ann. Phys. \textbf{253}, 55-82 (1997).
\bibitem{1997Arvind}
Arvind, K. S. Mallesh, and N. Mukunda, \emph{A generalized
Pancharatnam geometric phase formula for three-level quantum
systems.} J. Phys. A: Math. Gen. \textbf{30}, 2417-2431 (1997).
\bibitem{1964Bargmann}
V. Bargmann, J. Math. Phys. \textbf{5} 862 (1964).
\bibitem{1993Mukunda}
N. Mukunda and R. Simon, \emph{Quantum Kinematic Approach to the
Geometric Phase. I. General Formalism.} Ann. Phys. \textbf{228},
205-268 (1993).

\bibitem{Cory120}
Cory, D. G., Price, M. D. \& Havel, T. F.  Nuclear magnetic
resonance spectroscopy: An experimentally accessible paradigm for
quantum computing. \emph{Phys. D} \textbf{120}, 82(1998).

\bibitem{2001Sander}
B. C. Sanders, H. de Guise, S. D. Bartlett, and W. Zhang,
\emph{Geometric Phase of Three-Level Systems in Interferometry.}
Phys. Rev. Lett. \textbf{86}, 369-372 (2001).
\bibitem{1994Reck}
M. Reck and A. Zeilinger, \emph{Experimental Realization of Any
discrete Unitary Operator.} Phys. Rev. Lett. \textbf{73}, 58-61
(1994).


\end{thebibliography}
\end{document}